# Fullerene-Based Transparent Solar Cells with Average Visible Transmission Exceeding 80%

*Ruiqian Meng[1,2], Qianqing Jiang[1,2] \*, Dianyi Liu[1,2] \**

1 Key Laboratory of 3D Micro/nano Fabrication and Characterization of Zhejiang Province, School of Engineering, Westlake University, 18 Shilongshan Road, Hangzhou 310024, Zhejiang Province, China.

2 Institute of Advanced Technology, Westlake Institute for Advanced Study, 18 Shilongshan Road, Hangzhou 310024, Zhejiang Province, China.

AUTHOR INFORMATION

**Corresponding Author**

\* Email: liudianyi@westlake.edu.cn; jiangqianqing@westlake.edu.cn

ABSTRACT

Transparent photovoltaic (TPV) devices have the great potential to apply as smart windows in the construction and agriculture field. The efficiencies of TPVs are growing up quickly in recent years and the champion efficiency even exceeds 10%. However, the transparency is still hard to further improved after the average visible transmission (AVT) achieved to 73%. Each component of the TPV devices will influent the transparency of the TPV. To date, the TPV with the AVT over 80% have not been reported yet. In this work, we describe the fullerene-based highly transparent solar cells. The CuSCN/$C_{60}$ heterojunction is used as the effective light absorber. By finely optimizing the thickness of fullerene films and introducing the highly transparent electrodes, the TPV exhibits the AVT up to 82% while the device efficiency is above 0.3%. This study affords a new avenue to construct highly transparent TPV device.

**TOC GRAPHICS**

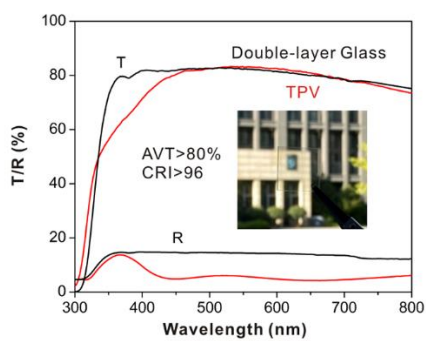

**Introduction**

Transparent photovoltaics (TPV) is a promising technique for building-integrated photovoltaics (BIPV) field.[1-3] The TPV devices can not only output the electricity power, but also allow the visible light passthrough which indicates the minimal impact on human vision.[4-7] Thus, the TPVs are potentially integrated as windows into buildings, vehicles, and agricultural greenhouses.[8] The previous research suggests that the semi-transparent BIPV installation can offset construction costs, which indicates the TPV development is benefits the environmental, energy and economic aspects.[9]

The research on TPV has made great progress in recent years.[10-14] The efficiency and transparency of TPV devices are promoted rapidly. The power conversion efficiency (PCE) is promoted to over 10% with an AVT of 52.9%.[15] Some other TPVs with high efficiency and transparency has also been reported. Xie *et al.* reported an organic transparent solar cell with all-near-infrared absorption materials. The device showed the PCE of 3.5% with the AVT of 61.5%.[16] Zuo *et al.* reported an organic-perovskite tandem transparent solar cell. The perovskite-based sub cell showed the PCE of 7.5% and the AVT of 68%, and the tandem cell exhibited the efficiency over 10% with the AVT of 52.9%.[15] These progresses greatly encourage the research on TPVs.

It is well-known that the high transparency is the unique property of TPVs different from opaque devices, however, the transparency of previous reported TPVs is still low and not satisfied for most of the commercial applications. To date, the highest AVT of transparent solar cells is reported up to 77.45%.[17] Though the transparency has already met the requirement of some special applications, it is still lower than 80% and not comparable with the window glass, which will result in some discomfort for the visual perception. There are two factors which mainly limit the development of high transparency of TPV device. The one is lacking in well-designed light absorption materials,[18-20] the other one is lacking in the highly transparent electrode.[3, 21-23] Here, we demonstrate fullerene-based highly transparent solar cells with AVT exceeding 80%. The CuSCN/$C_{60}$ heterojunction is used to absorb the light, and the highly conductive polymer/silver nanowire hybrid film is adopted as the top transparent electrode. The high transparency of TPV devices is comparable with the window glass after finely modified the fullerene film thickness.

**Results and Discussion**

Fullerenes have been widely used in organic photovoltaic solar cells. Recently, it is realized that fullerene and its derivatives not only play a crucial role as the electron acceptor and electron transport material in solar cells, but also make a critical difference in photon absorption and photocurrent generation.[24-25] Sit et al. prepared a CuSCN/$PC_{70}BM$ heterojunction solar cell, the PCE can reach over 6%.[26] Since CuSCN is a wide bandgap (3.49 eV) semiconductor material and it is almost transparent for the light, the $PC_{70}BM$ with the bandgap of 1.7 eV is the main light absorber in the solar cell.[27] The previous studies suggest that fullerenes are excellent ambipolar semiconductors which ensures produce the high performance of fullerene-based solar cells. Here, we form the CuSCN/$C_{60}$ heterojunction structure and use $C_{60}$ as the light absorption material, since $C_{60}$ has the similar absorption ability with $PC_{71}BM$ and the film thickness can be easily controlled by thermal evaporation method.

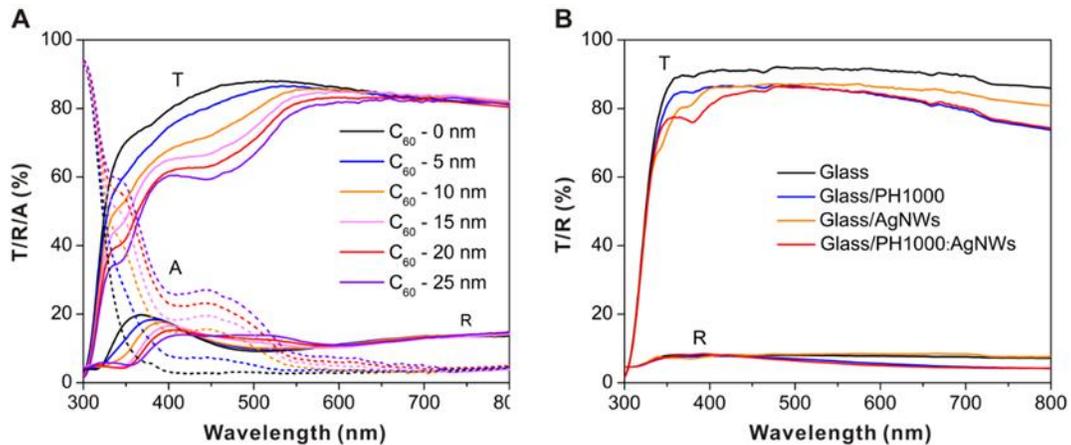

**Figure 1.** Optical properties of devices and transparent electrodes. (A) The transmission (T), reflection (R) and absorption (A = 1 – T – R) spectra of ITO/PEDOT/CuSCN/$C_{60}$/BCP devices with various thicknesses of $C_{60}$ films (from 0 to 25 nm). (B) The transmission and reflection spectra of the different electrode materials on glass substrates.

The specific device with the structure of ITO/PEDOT/CuSCN/$C_{60}$/BCP is used to investigate the transmission property of the expected TPV device before the introduction of the electrode. The devices with $C_{60}$ films of different thicknesses from 0 to 25 nm are prepared in this study. As illustrated in Figure 1A and Table 1, as the thickness of the $C_{60}$ film increased (0-25 nm), the absorption of the corresponding device is gradually increased. Meanwhile, the reflection at the visible light region is also slightly increased, which results in a significant decrease in the transmission. Moreover, most of the devices exhibit the AVT greater than 80%, and device with the $C_{60}$-5 nm film shows the highest AVT of 85.6%. When the thickness of $C_{60}$ film reaches 20 nm, the AVT of the device reaches the critical boundary of 80%. According to the above results, the thicknesses of $C_{60}$ films in TPV devices are limited to lower than 20 nm in the following experiments, which is expected to get the high transparent solar cells with the AVT > 80%.

**Table 1.** AVT and CRI of ITO/PEDOT/CuSCN/$C_{60}$/BCP devices with various thicknesses of $C_{60}$ films.

| Thickness (nm) | 0 | 5 | 10 | 15 | 20 | 25 |
|---|---|---|---|---|---|---|
| AVT (%) | 86.9 | 85.6 | 84.0 | 82.1 | 79.9 | 77.6 |
| CRI | 98.7 | 96.9 | 96.8 | 94.9 | 94.0 | 86.2 |

Transparent electrode is another key factor for TPVs.[28-30] Besides employing the highly transparent ITO glass as the electrode, a highly conductive polymer-based film is also be introduced as the transparent electrode. PH1000 is a kind of highly conductive conjugating-based polymer that is broadly used as the transparent electrode in organic photovoltaic field.[31-32] Using the lamination transfer technique, PH1000 could be easily transferred to the device surface and act as the top transparent electrode.[33-35] To further reduce the sheet resistance, the silver nanowires (AgNWs) solution with the concentration of 3 mg/ml is added into PH1000 precursor in this study. The results of film resistant test show that the resistance of PH1000: AgNWs hybrid film is reduced from 141±31 Ω/sq for the original PH1000 film to 84±22 Ω/sq. Meanwhile, the transmission spectra have no obvious change (Figure 1B). The AVT of PH1000 film and PH1000: AgNWs hybrid film show the close value of 84.9% and 85.0% respectively, which indicates that the AgNWs additive has little effect on the film transparency.

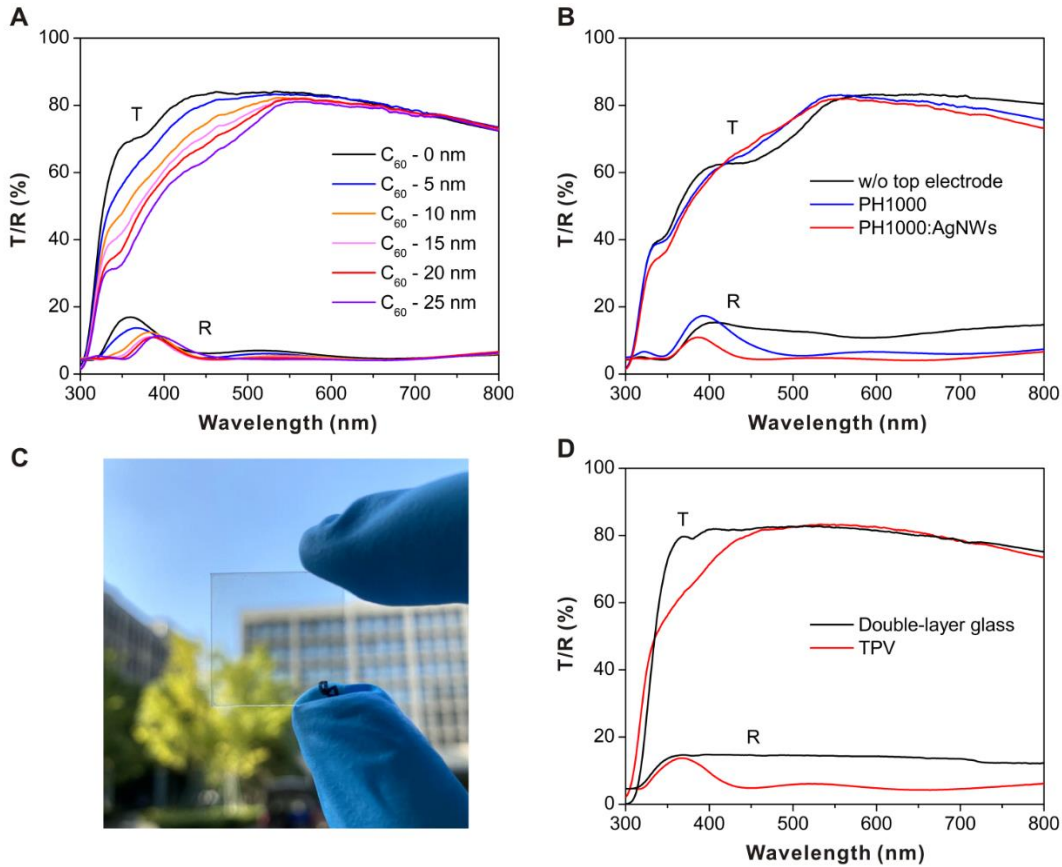

**Figure 2.** Optical properties of the TPV devices. (A) Transmission spectra of TPV devices with the $C_{60}$ films thicknesses from 0 to 25 nm. (B) The transmission and reflection spectra for the corresponding $C_{60}$ - 20 nm TPV devices with different top electrodes. (C) Photograph of $C_{60}$ - 20 nm TPV device. (D) The transmission and reflection spectra of double-layer glass and $C_{60}$ - 5 nm TPV device.

To acquire more information of the PH1000 and PH1000: AgNWs electrodes, scanning electron microscope images (SEM) of related films were shown in Figure S1. It can be found that the randomly cross-linked AgNWs were built with low coverage ratio, therefore, it may favor the transmission of incident light.[36] The high conductivity and transparency indicate that the PH1000: AgNWs hybrid film is a good candidate for TPV transparent electrode.

According to the above research, the PH1000: AgNWs hybrid film is subsequently deposited onto the surface of ITO/PEDOT/CuSCN/$C_{60}$/BCP device to fabricate the TPV device by transfer lamination method. The schematic diagram of electrode transfer lamination is illustrated in Figure S2. Firstly, the mixture of PH1000: AgNWs is deposited on the PDMS substrate. After the film dried in air, the PDMS substrate is put on top of the ITO/PEDOT/CuSCN/$C_{60}$/BCP device to make the PH1000: AgNWs stick with the bottom device. Then the PDMS substrate is peeled off and the PH1000: AgNWs is left as the top electrode. The top-view SEMs of the TPV device are shown in Figure S3. The staggered arrangement of AgNWs on the organic layer can be observed clearly, and the morphology of PH1000: AgNWs electrode has not been changed after transferred. The results of TPV optical measurement are summarized in Figure 2A and Table 2. It showed that when the $C_{60}$ is no more than 20 nm in thickness, all of the TPV devices exhibit the AVT greater than 80%. In addition, based on the $C_{60}$ films with the thickness of 20 nm and 25 nm, the AVTs of corresponding TPV devices are even higher than the device before the top electrode was added. The transmission (T) and reflection (R) characterizations for the corresponding $C_{60}$-20 nm devices are performed separately in Figure 2B. After the PH1000: AgNWs electrode lamination, the transmission of TPV device in the range of 420 nm to 590 nm is higher than the device without the top electrode and result in a certain improvement of the AVT of TPV. We infer that the transmission improvement is stems from the anti-reflection effect of the PH1000: AgNWs hybrid electrode. Figure 2B shows that the reflection of TPV device is significantly lower than the device without the top electrode, which suggests that more photons can go into/through the TPV device. The anti-reflection effect can offset the photons loss by electrode absorption and result in higher transmission. It suggests that the reflection is mainly determined by the surface layer which is the hybrid PH1000: AgNWs electrode in the TPV devices.[37-39]

**Table 2.** AVT and CRI of TPV devices with various thicknesses of $C_{60}$ films.

| Thickness (nm) | 0 | 5 | 10 | 15 | 20 | 25 |
|---|---|---|---|---|---|---|
| AVT (%) | 83.2 | 82.6 | 81.2 | 80.4 | 80.2 | 78.9 |
| CRI | 98.8 | 98.2 | 95.4 | 93.6 | 96.8 | 95.7 |

It is notable that the color rendering index (CRI) of TPV devices is up to 98, which indicates the devices are almost colorless and show the minimum impact on vision.[11, 13] As shown in the photograph of the CuSCN/$C_{60}$ heterojunction TPV cell with $C_{60}$ thickness of 20 nm in Figure 2C, the device looks clear and substantially transparent. Since the double-layer glass is one of the most popular materials for the building and vehicle windows, the double-layer glass is also be used as the reference to descript the transparent property of the highly transparent TPVs here. The transmission and reflection spectra of double-layer glass and $C_{60}$ - 5 nm TPVs are shown in Figure 2D. Since the CuSCN/$C_{60}$ heterojunction TPV devices present the AVT of 80.2% - 82.6% and CRI of 93 - 98, which are comparable with double-layer glass (AVT of 82% and CRI of 99.1), the highly transparent TPVs essentially match the optical requirements for the application on window glass in the architecture field.

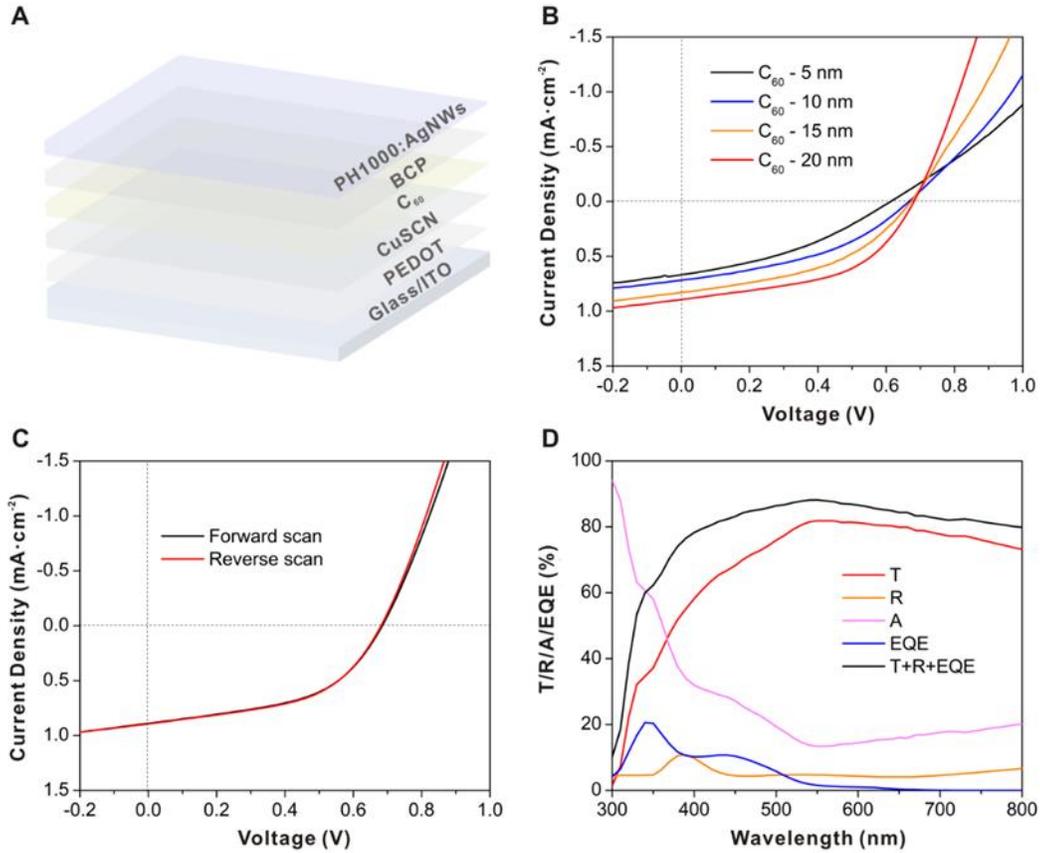

**Figure 3.** Performance characteristics of TPV devices. (A) Architecture of the TPV device. (B) J-V characteristics for TPV devices with the different thicknesses of $C_{60}$. (C) J-V curves from forward scan and reverse scan for the $C_{60}$-20 nm TPV device. (D) Transmission (T), reflection (R), absorption (A) and external quantum efficiency (EQE) properties for the corresponding $C_{60}$-20 nm TPV device.

The energy conversion performance is also tested for CuSCN/$C_{60}$ heterojunction TPVs with AVT > 80%. The similar device architecture is shown in Figure 3A. *J-V* characteristics of the device is measured under AM 1.5G illumination condition. The *J-V* curves and the key performance parameters are summarized in Figure 3B, Figure S4 and Table S2. Distinctly, the PCE exhibits an upward trend with the increased thickness of $C_{60}$ (Table S2). This result largely depends on the light absorption capacity of the active layer, which can also be confirmed in the transmission curve of the film (Figure 1A). The PCE of the champion TPV is up to 0.31%, with the open-circuit voltage ($V_{OC}$) of 0.680 V, the short-circuit current density ($J_{SC}$) of 0.893 mA cm$^{-2}$ and the fill factor (FF) of 50.3%. Meanwhile, the forward scan and reverse scan curves of the device basically coincide, which is serviceable in proving the well-fabrication of the device (Figure 3C). In addition, the integrated $J_{SC}$ by the external quantum efficiency (EQE) test is 0.85 mA cm$^{-2}$, which matches well with the $J_{SC}$ obtained by the *J-V* measurement, and further verifies the performance of the device. To satisfy the photon balance consistency check,[40] the optical properties of the TPV devices are systematically researched. As shown in Figure 3D, the T, R, A and EQE curves of the complete device are summarized. It can be proved that EQE + T + R < 100% in all spectral ranges.

**Conclusion**

We demonstrate the CuSCN/$C_{60}$ heterojunction TPV devices with the PH1000: AgNWs hybrid top electrode. AVT > 80%, CRI > 93%, accompanying with the PCE of up to 0.3% are achieved with the TPV devices. The transparent performance of TPVs is even comparable with the double-layer glass which is broadly used as the window in the architecture field. The study provides a novel avenue for the fabrication of high-transparency TPV devices, promotes the further development of TPVs and smart windows/buildings. The reported TPV cells in recent years with over 50% of AVT are summarized in Figure S5 and Table S1. The research on highly transparent TPVs is increasing dramatically, and device transparency is growing rapidly from ~60% to over 80% in recent three years.

The transparency limitation of TPV is expected to be solved by the growing researches in the near future. However, the visible light absorption is intentionally limited to achieve the high transparency of device. The photocurrent is then quite low which results in the PCE of only 0.3% in the device. To improve the efficiency of the highly transparent TPVs in future researches, one of the key strategies is to develop the non-visible-light selective absorption materials and increase the non-visible light absorption capability. The initial demonstration of AVT over 80% for TPVs is encouraged to power smart buildings and meet the critically high level of aesthetics requirements for actual applications.

ASSOCIATED CONTENT

**Supporting Information**.

Experimental details of device fabrication and characterization. Top-view SEM, schematic diagram, statistical graphs and summary of device parameters, summary of reported TPV cells with over 50% of AVT.

AUTHOR INFORMATION

**Corresponding Authors**

**Dianyi Liu** - Email: liudianyi@westlake.edu.cn

Qianqing Jiang- Email: jiangqianqing@westlake.edu.cn

**Notes**

R. M., Q. J. and D.L. have filed a patent application based on the work in this manuscript.

ACKNOWLEDGMENT

**General:** We thank instrumentation and service center for physical sciences for the facility support and technical assistance.

**Funding:** This work was supported by the Westlake Education Foundation.